\documentstyle[11pt,twoside,jltp,epsfig]{article}

\title{The Long Range  Interaction and the Relaxation
in Glasses at  Low Temperatures}
\author{A.L. Burin\address{Department of Chemistry, Tulane University, New 
Orleans,
LA 70118, USA} and I.Ya.Polishchuk\address{RRC Kurchatov Institute, Kurchatov 
Sq. 1, 123182 Moscow,
Russia}\address{Max-Planck-Institut f\"{u}r Physik Komplexer Systeme, 
D-01187 Dresden, Germany}}

\runninghead{Alexander L. Burin and Il'ya Ya. Polishchuk}{The Long Range 
Interaction and the Relaxation
in Glasses at Low Temperatures}

\begin{document}

\begin{abstract}
We describe the interaction stimulated relaxation in the ensemble of 
two-level systems, responsible for low temperature kinetics and thermodynamics
properties of amorphous solids. This relaxation gets significant at sufficiently 
low
temperature when phonons are substantially frozen out. We show that in the 
realistic experimental
situation the measuring field strongly accelerates the interaction
stimulated relaxation. The characteristic temperature and field dependences
of the relaxation rate are found when the rate is affected both by the interaction between two
level systems and by the external field.

\end{abstract}

\maketitle

\section{Introduction}

For a long time the standard model of \textit{non-interacting}
two-level-systems (\textit{TLS}) \cite{ahvp} has served as a good background for
understanding experimental data in glasses at low temperatures 
$T\leq 1K$. The further investigations revealed that below $100mK$ almost all measurements
in dielectric glasses dealing with their relaxation properties 
\cite{b11-01,b11-011,b11,b11-1,b11-2} cannot be treated ignoring the \textit{TLS} interaction. 
For this reason one can suppose that that below $100mK$
\textit{TLS} manifest the collective behavior induced by the 
interaction between them.

It is well established that in amorphous solids  \textit{TLS} are coupled with
phonons. In dielectric glasses at $T\ll \Theta_{D}$ ($\Theta_{D}$ is
the Debye temperature) coupling with acoustic phonons plays a main role.
The virtual exchange of  \textit{TLS} by acoustic phonons results in the 
indirect interaction between  \textit{TLS}, which decays with the distance $R$ as $R^{-3}$. In particular, this interaction gives rise to the flip-flop transitions between 
two  \textit{TLS} (see Fig. 1). The transition amplitudes also decay with 
the distance as $R^{-3}$ \cite{Attempts,b7}. In a number of dielectric 
glasses,  \textit{TLS} possess their own dipole moment. In this case the electrostatic dipole-dipole interaction can become the dominating interaction between \textit{TLS}.\cite{Hunklinger} 
This interaction also decreases with the distance as $R^{-3}$. It was 
experimentally discovered by Arnold and Hunklinger \cite{hun1} ( see also 
\cite{Hunklinger})
that the $1/R^{3}$ interaction contributes to the spectral diffusion of  the 
\textit{TLS} energy. 
This spectral
diffusion involves the dynamic fluctuation of  the \textit{TLS} energy due to 
its interaction with neighboring  \textit{TLS}, making transitions between 
their levels \cite{Halperin 1968}. 

Along with the interaction significance for the acoustic \cite{hun1,Hunklinger} and optical \cite{v,h,j} 
hole burning experiments, it also affects the  decay
of coherent echoes \cite{gra}. Also the interaction between  \textit{TLS} has 
been
revealed in the non-equilibrium dielectric measurements \cite{b11-1,osher1}, 
giving rise to the reduction of the spectral density near zero energy. This is interpreted as the dipole gap formation \cite{b95}.  
In addition, it has been found
that the transition from coherent to incoherent tunneling takes place if the
typical interaction energy exceeds tunneling splitting \cite{hun97}.
Recently discovered anomalous low-temperature behavior of amorphous solids in 
the external magentic field \cite{m1,m2,m3,m4,m5,m6} can also be associated with the  \textit{TLS} interaction.

The relaxation in a disordered system can be due
to delocalized excitations. The single particle energy delocalization is not efficient 
for the ensemble of interacting  \textit{TLS} because the static energy disordering is very 
strong there. Therefore, the Anderson localization \cite{classical} of all 
excitations takes place. In other words, 
the localization in the disordered  \textit{TLS} system is due to a large energy 
level mismatch for a typical pair of  \textit{TLS} (Fig. 1A) compared with the 
flip-flop transition amplitude for this pair. Suppose that a  \textit{TLS}
excitation energy can experience large time-dependent fluctuations. These
fluctuations can reduce the energy mismatch value, stimulating level crossing and 
supporting flip-flop transitions (see the resonant pair in Fig. 1B). Such 
fluctuations can be induced by either  the external alternating (measuring) 
field or they can be due to the spectral diffusion, caused by transitions of 
neighboring \textit{TLS}.  Thus, such fluctuations 
can promote the delocalization in the system. 
In the absence of the external field this delocalization can be treated 
as a self-consistent process in which the spectral diffusion induces \textit{TLS} transitions. In its turn, \textit{TLS} transitions cause the spectral diffusion due to \textit{TLS} interaction. 

This paper  is a comprehensive
presentation of our results concerning the effect of $R^{-3}$ interaction
on the dynamic properties of amorphous solids. 
It is organized as follows. In Sect. 2 we address the single particle 
localization problem for zero temperature and zero external alternating 
field. 
In Sect. 3 the effect of an external alternating field on the single particle 
localization is studied. We demonstrate that the field having a sufficiently 
large amplitude and a small frequency can break the Anderson localization. In 
Sect. 4 we study the finite temperature many-body delocalization of a 
\textit{TLS} energy within the  \textit{ TLS} ensemble induced by the self-
consistent dynamics of resonant pairs (see Fig. 1B). Sect. 5 is devoted to the 
most complicated, but experimentally relevant regime where both external field 
and interaction of  \textit{TLS} must be taken into account simultaneously. In summary we present the brief discussion of the application of our theory to the 
experimental studies.

\section{Anderson localization of  the \textit{TLS} energy}

Following Ref.~\cite{ahvp}, we accept that the distribution function for 
the bias asymmetries $\Delta $ and tunneling amplitudes $\Delta _{0}$ (see Fig. 1) obeys 
the
universal distribution 
\begin{equation}
P(\Delta ,\Delta _{0})=\frac{P}{\Delta _{0}}.  \label{eq:AHWP}
\end{equation}
An isolated  \textit{TLS} can be described by the standard pseudospin 
Hamiltonian 
\begin{equation}
h_{i}=-\Delta _{i}S_{i}^{z}-\Delta _{0i}S_{i}^{x}.  \label{eq:spin}
\end{equation}
The interaction between  \textit{TLS} can be expressed as\cite{Hunklinger,b6}
\begin{equation}
\widehat{V}=\frac{1}{2}\sum_{i,j}U(R_{ij})S_{i}^{z}S_{j}^{z},\hspace{3mm}
U(R_{ij})=\frac{U}{R_{ij}^{3}},  \label{eq:int}
\end{equation}
where $R_{ij}$ is the distance between two  \textit{TLS} and $U$ is the
characteristic coupling constant.

Consider a relaxation of a sole excited  \textit{TLS} with certain energy
splitting $E=(\Delta ^{2}+\Delta _{0}^{2})^{\frac{1}{2}}$. The 
relaxation channel independent of phonons is hopping of
an excitation from excited  \textit{TLS} to another  \textit{TLS} with 
parameters $\Delta
^{\prime },\Delta _{0}^{\prime }, 
E^{\prime}=(\Delta^{\prime 2}+\Delta _{0}^{\prime 2})^{\frac{1}{2}}$, 
which is initially in its  ground 
state (see Fig. 1B). As a result, first  \textit{TLS}  goes into
the ground state, while second  \textit{TLS} becomes excited. The
inverse process also takes place so that the pair of  \textit{TLS} under 
consideration
can be detected in one of the two states separated by the small energy 
$\mid E-E^{\prime }\mid $. In what follows, these two states of a  \textit{TLS} 
pair
will be referred to as a flip-flop configuration. Such a  \textit{TLS} pair
can be considered as a new type of the two-level system with the
asymmetry $\Delta _{p}=\mid E-E^{\prime }\mid $. In addition, as follows
from Eqs. (\ref{eq:spin}), (\ref{eq:int}), the tunneling amplitude 
coupling two states of the flip-flop configuration of a  \textit{TLS} pair is 
given by
the relation \cite{b7,Laikhtman 1985,iyp}
\begin{equation}
\Delta _{0p}(R)\approx U(R)\frac{\Delta _{0}\Delta _{0}^{\prime }}{%
EE^{\prime }}.  \label{eq:transampl}
\end{equation}

Thus one can describe the evolution of initially excited  \textit{TLS} in terms of
the transition between the two states of the flip-flop configuration of a
 \textit{TLS} pair (Fig. 1B).
If a certain two-level system with the parameters $\Delta$ and $\Delta_{0}$
was initially, say, in the first state, the time-dependent probability 
$W_{2}(t)$ to
find the system in the second state is given by the solution of the
corresponding time-dependent Schr\"{o}dinger equation (hereafter we set 
$\hbar=k_{B} =1$) 
\begin{equation}
W_{2}(t)=\frac{\Delta_{0}^{2}}{\Delta^{2}+\Delta_{0}^{2}}\sin^{2}(t \sqrt{%
(\Delta^{2}+\Delta_{0}^{2})}).  \label{flugge}
\end{equation}
This relation applied to the flip-flop transition within the  \textit{TLS} pair 
(see Fig. 1) means that the transition efficiency is
noticeable provided the parameters of a  \textit{TLS} pair obeying the 
\underline{resonance} condition
\begin{equation}
\Delta_{p}=\mid E-E^{\prime}\mid<\Delta_{0p}(R),  \label{eq:reson}
\end{equation}
where the pair transition amplitude $\Delta_{0p}$ is defined by Eq. 
(\ref{eq:transampl}). Hereafter such a  \textit{TLS} pair is referred to as a 
\textit{resonant} \textit{pair(RP)}.
Then, the frequency of quantum mechanical oscillations of population between the
potential wells of the flip-flop configuration given by (cf. Fig. 1B, Eq. 
(\ref{flugge})) 
\begin{equation}
\tau^{-1}=\sqrt{(\Delta_{p}^{2}+\Delta_{0p}^{2})}\approx\Delta_{0p}(R)
\label{tau}
\end{equation}
is completely determined by the pair flip-flop transition amplitude  
$\Delta_{0p}(R)$ (\ref{eq:transampl}). The inverse of this frequency defines the transition time $\tau$. 


\begin{figure}[tbp]
    \centering
    \includegraphics[width=5in,clip]{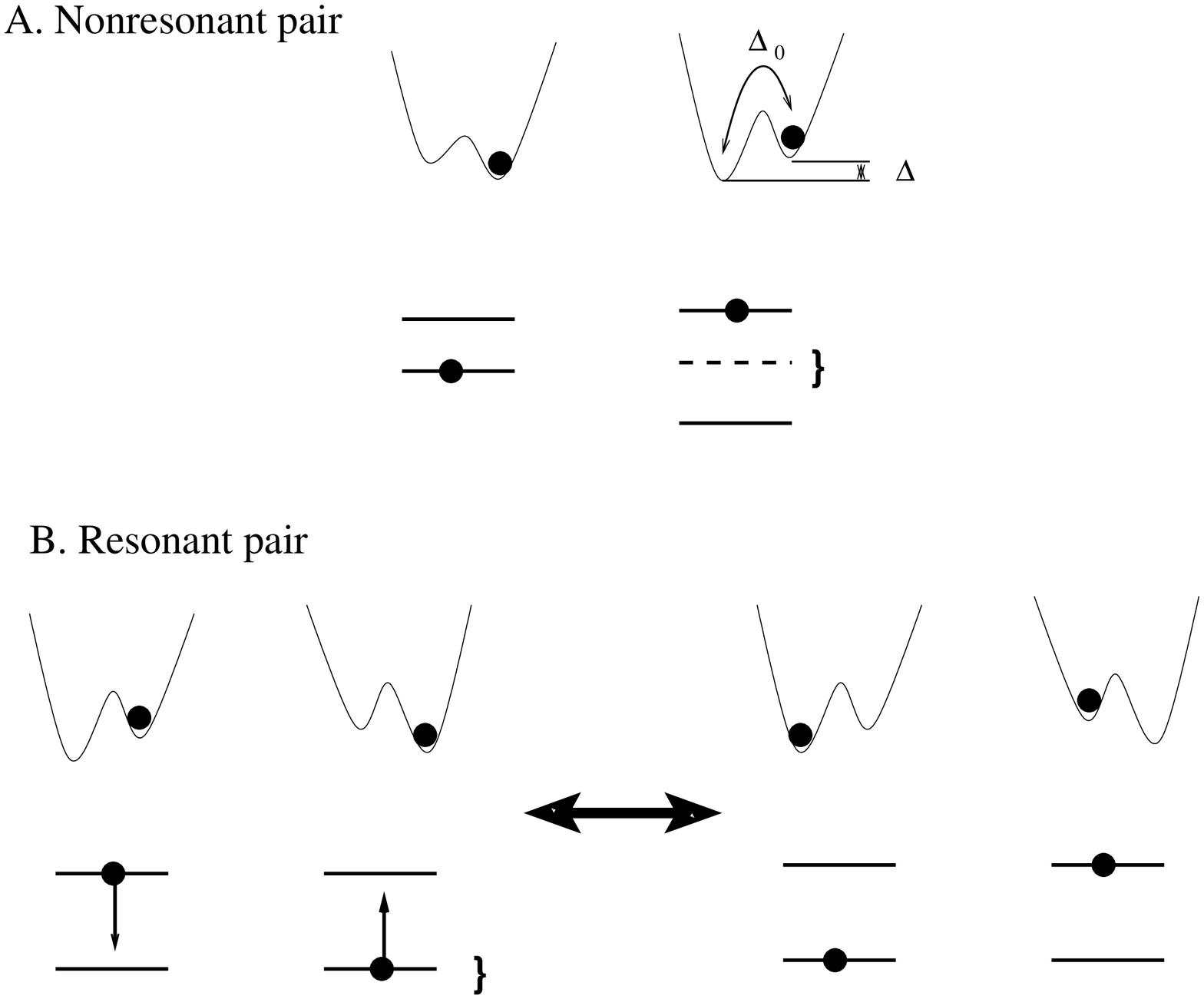}
    \caption{\label{Res1}
Energy transport between excited  \textit{TLS} (left) and nonexcited  
\textit{TLS} (right). It is not efficient in the off-resonant case (A), where 
the energy level mismatch exceeds the resonant interval indicated by the figure 
bracket.  For the resonant pair (B) the interaction of  \textit{TLS} induces 
flip-flop transitions mixing up two possible quasi-degenerate states, where 
either first or second  \textit{TLS} is excited, while the remaining TLS is in 
its ground state.}
\end{figure}%

The Anderson delocalization and/or localization are explicitly associated with 
a  \textit{TLS} capability to form resonant pairs. If the typical  \textit{TLS} 
has on 
average the large 
number of resonant neighbors, one would expect the delocalization. In fact, each 
resonance shares the energy between two  \textit{TLS}. One can use a percolation 
analogy, connecting all resonant pairs into a cluster. When the infinite cluster 
can be formed, excitations become delocalized (cf. Ref. \cite{ShklovskiiBook}). It 
is clear that we need to have more than one resonant neighbour for each  
\textit{TLS} ($2$-$3$ should determine the threshold \cite{ShklovskiiBook}) to make 
such a cluster. When the average number $N$ of resonant pairs formed by one  
\textit{TLS} is much less than unity, this TLS cannot share its energy with 
anybody (Fig. 1A) or it has one resonant neighbour (Fig. 1B), with whom it hybridyzes its excitation. The probability to add more resonances decreases by a 
small factor of $N\ll 1$, so all excitations are localized.

How many resonances does a given excited  \textit{TLS} with the parameters 
$\Delta_{0},E$ form with other  \textit{TLS}? 
To answer this question, one should calculate the average number 
$N[\Delta_{0},E]$
of  \textit{TLS}, forming a resonant pair with selected excited  
\textit{TLS}. 
Using Eqs. (\ref{eq:AHWP}), (\ref{eq:transampl}), (\ref{eq:reson}), we find 
\begin{eqnarray}
N[\Delta_{0},E]
\nonumber\\
=P\int d\Delta^{\prime}\int\frac{d\Delta_{0}^{\prime}}{
\Delta_{0}^{\prime}}\int_{a}^{L}d^{3}\bf{r}
\cdot  \Theta(\Delta_{0p}(r)-\Delta_{p})\Theta(r-a)\Theta(L-r) 
\nonumber\\
\approx\frac{\Delta
_{0}}{E}\chi\ln\left( \frac{L}{a}\right), 
\nonumber\\
\chi=4\pi PU. \label{eq:N0}
\end{eqnarray}
Here $\Theta$ is the Heaviside function involving the resonant condition
Eq.(\ref{eq:reson}) and minimum and maximum distances between interacting TLS 
($\Theta(x)=1$, $x\geq 1$, $\Theta(x)=0$, $x<0$),  
$L$ is the size of the whole system and $a$ is the minimum 
distance between two
 \textit{TLS}. Since in all glasses the parameter $\chi \propto PU$ is very 
small (see e. g. \cite{Freeman,b6})
\begin{equation}
\chi = 4\pi PU \leq 10^{-3}-10^{-2}, 
\label{eq:chi}
\end{equation} 
the number of resonant neighbors is much less than unity for any reasonable 
sample size $L$. Therefore only a small fraction of two level systems belongs to 
resonant pairs, while majority of the others are immobile. This proves the full 
Anderson localization of all excitations in this system.

\section{Delocalization of Floquet states}

Consider the effect of an external alternating field on the
Anderson localization within the  \textit{TLS} ensemble. This problem is 
experimentally important because almost all mesuring techniques use the external 
alternating field to probe the system response. At ultralow temperatures 
$T<0.1K$ it is very difficult to make the field small enough in order to avoid its nonlinear effect. It turns out that the interaction-induced relaxation 
is also affected by the field. In this subsection we analyze the simplified 
single-particle problem and a more realistic case will be treated later in Sect. 
5 of this paper.
 
An approximate zero
temperature condition means that one can neglect the interaction between
excited  \textit{TLS} because their total number is negligibly small in this case. 
In particular, there is no spectral diffusion. Consider the
effect of the alternating field on the energy spectrum of excitations.\cite{bkp2001} 
Due to the field, the energy splitting $\Delta _{i}$
acquires the oscillating part $a_{i}cos(\omega t)$. Here $a_{i}$ is a
coupling energy between  \textit{TLS} in the site $i$\ and the external field, 
that is
a product of the field strength (electric field $E$ or elastic stress $%
\varepsilon $) and the coupling constant (\textit{TLS} dipole moment $\mu $ or 
strain
tensor $\gamma $, respectively \cite{Hunklinger}). We assume that the field 
varies
sufficiently slow and the amplitude $a_{i}$ is small
compared to the typical scale of the excitation energy $E_{i}=\sqrt{\Delta
_{i}^{2}+\Delta _{0i}^{2}}$ 
\begin{equation}
\omega \ll a_{i}\ll E_{i}.  \label{eq:main_cond}
\end{equation}%
The assumption of a small field amplitude is needed in order to treat 
the field as a weak perturbation. The assumption of low 
frequency is satisfied in the vast majority of acoustic and dielectric 
experiments in glasses, because the minimum field amplitude usually exceeds 
$10^{-4}K$ while the maximum frequency is usually below $100kHz$ corresponding to the 
energy $10^{-5}K$. Moreover, usually a ratio $a/\omega$ exceeds unity by 
several orders of magnitude. 

\begin{figure}[tbp]
    \centering
    \includegraphics[width=5in,clip]{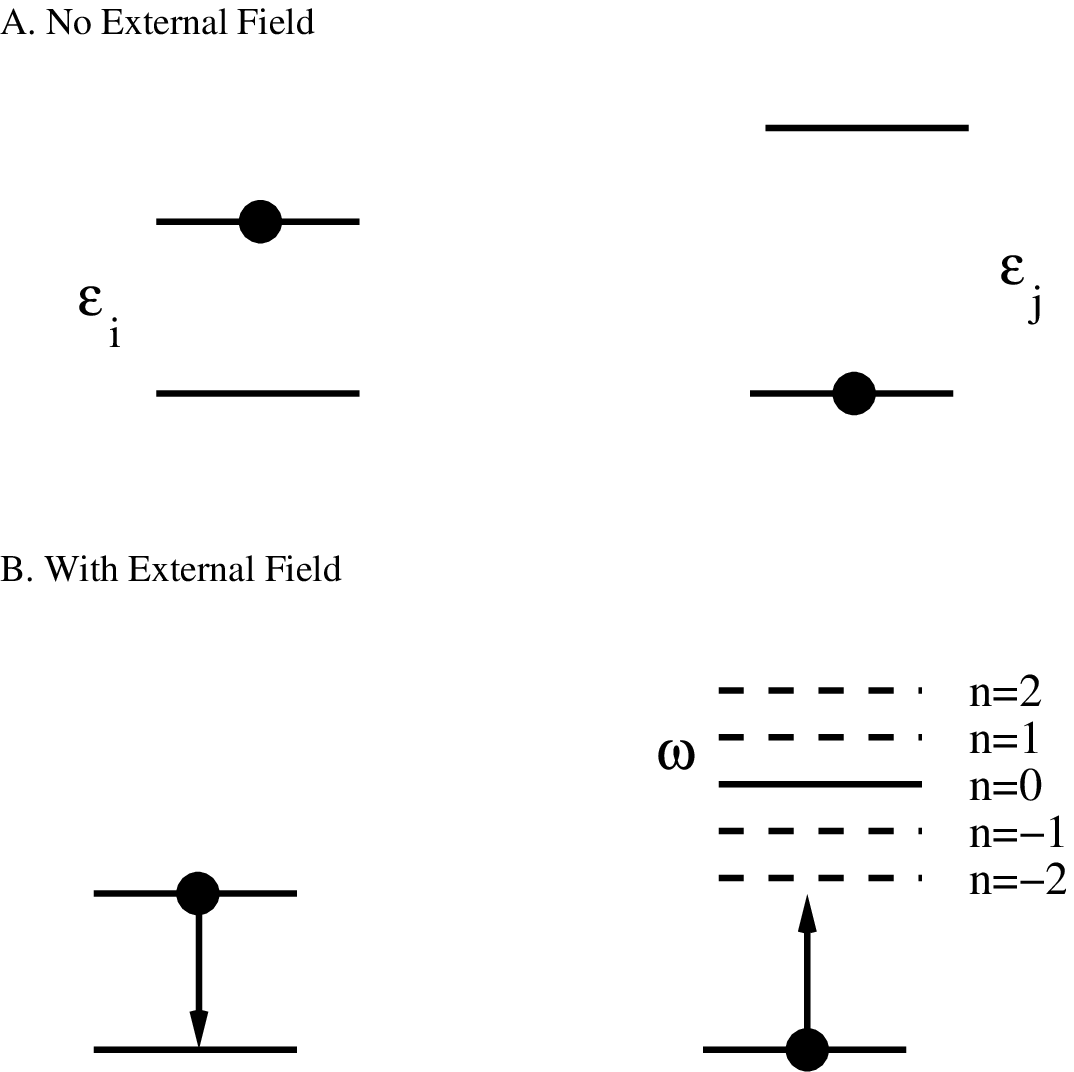}
    \caption{\label{Res3}
The formation of a resonant pair from the initially non-resonant one (A) in the 
external alternating field. The coherent excitation dressing (B) enables transitions 
with the energy change given by the integer number $n$ of field frequences. The 
case $n=-2$ satisfies the resonance conditions 
( see Eq. (\ref{ander})).}
\end{figure}%

Let $b_{i}$ be the amplitude of the excited  \textit{TLS} state at the site $i$. 
An evolution of a  single-particle excitation can be described by the Schr\"{o}dinger
equation with the external field involved 
\begin{equation}
i\dot{b}_{i}=\left( E_{i}-a_{i}\cos \left( \omega t\right) \right)
b_{i}+\sum_{j}U_{ij}b_{j}.  \label{efse2}
\end{equation}%
\noindent Here $U_{ij}$ are hopping amplitudes (\ref{eq:transampl})
between pairs of sites $i$ and $j$. The subsystem of \textquotedblleft
active\textquotedblright\ or resonant  \textit{TLS} effectively responsible for 
the
dynamics consists of  \textit{TLS} with $\Delta _{i}\leq \Delta _{0i}$. For this
subsystem, the factor $\frac{\Delta _{0i}\Delta _{0j}}{E_{i}E_{j}}$ is close
to unity and can be omitted. Then, considering only the subsystem of
\textquotedblright active\textquotedblright\  \textit{TLS}, we put\vspace{-
0.2cm} 
\begin{equation}
\left\vert U_{ij}\right\vert =U/R_{ij}^{3}  \label{b32}
\end{equation}
and assume the uniform density for the
energy levels $P\left( E \right) =P$.\cite{b6}

Equation (\ref{efse2}) can be rewritten in terms of the Floquet state amplitude 
$d_{in}$ (see Appendix) 
\begin{eqnarray}
\varepsilon d_{in} =\left( E_{i}-n\omega \right)
d_{in}+\sum_{jp}T_{in;jp}d_{jp},  
\hspace{0.5cm}
T_{in;jp} =U_{ij}J_{p-n}\left( \frac{a_{j}-a_{i}}{\omega }\right)
\label{renampl}
\end{eqnarray}
with $\varepsilon $ being an eigenvalue of a Floquet eigenstate quasi-energy. 
Here $J_{p}(x)$ is the Bessel function of order $p$.

Equation (\ref{renampl}) resembles those usually considered in studying the
Anderson localization when the disorder is static. If the hopping term $
T_{in;jp}$ is neglected, the eigenstates of the system in the external field
correspond to localized single-site excitations of a  \textit{TLS} coherently 
dressed
by $n$ \textquotedblright quanta\textquotedblright\ of the frequency $\omega$
its position and the number of the dressing quanta. Therefore, the problem is
reduced to investigating the delocalization of these dressed excitations 
(\textit{DE}).

Suppose that an excitation is created at the site $i$. It can be described as a 
\textit{DE} with certain double indices $i0$. To leave the state $%
i0$ for another \textit{DE} state $jn$, the condition of resonant coupling
should be satisfied, as shown in Fig. 2,
\begin{equation}
\left\vert E_{i}-n\omega -E_{j}\right\vert \leq \left\vert
T_{i0;jn}\right\vert =\left\vert U_{ij}J_{n}\left( \frac{a_{j}-a_{i}}{\omega 
}\right) \right\vert .  \label{ander}
\end{equation}
This criterion is similar to multiphoton resonances 
in the nonlinear optics. The successful transition between two states can 
be accompanied by an absorption or an emission of $n$ quanta,  
to provide the energy conservation.  
These processes are efficient, when the field amplitude is large to 
ensure the strong nopnlinearity like in our case. 

The local amplitudes $a_{i}$ vary from site to site due to
fluctuations in coupling constants. Let us denote the average of a 
difference $\left\vert a_{i}-a_{j}\right\vert $ by $a$ having the same order
of magnitude as amplitudes $a_{i}$.

The argument of the Bessel function in Eq. (\ref{renampl}) is a large parameter
of order of $a/\omega\gg 1$ (see Eq. (\ref{eq:main_cond})). 
If $\left\vert n\right\vert >a/\omega$, the
magnitude of the corresponding Bessel function is exponentially small,
entailing a negligible probability of resonant coupling. The opposite
condition $a>\mid n\mid\omega$ means that, in order to have resonant
coupling, the energy difference should be at least less than the field amplitude 
$a$:
\begin{equation}
\mid n\mid<a/\omega\rightarrow\hspace{2mm}\left\vert E_{i}-E_{j}\right\vert
<a.  \label{smallomega}
\end{equation}
On the classical language this condition means that the interaction of TLS 
can be efficient only when the external field stimulates their 
real level crossing. 

Under condition (\ref{smallomega}) and $a/\omega \gg 1$ the Bessel function
in (\ref{renampl}) can be approximately replaced by its asymptotic value. Omitting the
standard oscillating prefactor which plays no role in a random discrete
problem, one can estimate the coupling amplitude for \textit{DE} Eq. (\ref{b32}) 
as 
\begin{equation}
\left\vert T_{i0;jn}\right\vert \approx \frac{U}{R_{ij}^{3}}\sqrt{\frac{
\omega }{a}},  \label{renampl1}
\end{equation}

Consider the delocalization of an excitation due to the
alternating field within the framework of the concept of resonant
coupling. Two sites $i$ and $j$ are in the resonance when the condition Eq. (\ref
{ander}) is satisfied at least for some $n<a/\omega $ (see Eq. 
(\ref{smallomega}) and Fig. 2). Since the energy splitting between two 
subsequent levels is
equal to the field frequency $\omega $, the resonance always occurs if the
hopping amplitude $T_{i0;jn}$ exceeds $\omega $ and condition (\ref%
{smallomega}) is fulfilled. As directly follows from Eq. (\ref{renampl1} ),
this happens when sites $i$ and $j$ are separated by a sufficiently small 
distance $R_{ij}$
\begin{equation}
R_{ij}<r_{\ast }=\left( U/\sqrt{a\omega }\right)^{1/3}.  \label{cross_rad}
\end{equation}
Thus all \textit{TLS} $j$ located closer than the crossover distance $r_{*}$ to 
\textit{TLS} $i$ under consideration   
with $\left\vert E_{i}-E_{j}\right\vert <a$ are resonantly coupled with each 
other. 
Taking into account that the number of such centers in a unit volume is equal to $Pa$, 
we find that 
the total number of the resonant neighbors within the sphere of 
a radius $r_{\ast }$ is given by  
\begin{equation}
W\left( r_{\ast }\right) =\frac{1}{3}\chi 
\sqrt{\frac{a}{\omega}}  \label{eq:sh_dist_alt_res}
\end{equation}
At larger distances $R>r_{\ast }$ the coupling
amplitude (\ref{renampl1}) is less than $\omega $. Then one should treat 
separately all $a/\omega$ possible resonances (see Fig. 2). Thus for the given 
pair the probability of resonance 
increases by the factor of $a/\omega$ due to the increase in the number of 
possible resonances. On the other hand, the probability of each resonance 
decreases by the factor $\sqrt{\omega/a}$ because of the reduction in the flip-
flop transition amplitude (see Eq. (\ref{renampl1})) compared to the zero-field 
case Eq. (\ref{eq:transampl}). 
As a result, the total number of resonances in each layer $r_{\ast 
}<r_{1}<R<r_{2}$
\begin{eqnarray}
W\left( r_{1},r_{2}\right) \approx P\int_{r_{1}}^{r_{2}}d^{3}R\left( 
\frac{U}{R^{3}}\sqrt{\frac{\omega}{a}}\frac{a}{\omega}
\right) =\chi _{\ast }\ln \frac{r_{2}}{r_{1}}  
\nonumber \\
\chi _{\ast } = 4\pi PU\sqrt{\frac{a}{\omega }},  \label{eq:l_dist_alt_res}
\end{eqnarray}
increases by the factor $\sqrt{a/\omega}$ as compared to the zero field case 
(\ref{eq:N0}). 
Thus, in the precence of the external
field the parameter $W\left( r_{1},r_{2}\right) $ increases by the large
factor $\sqrt{a/\omega }$.

According to the Levitov's analysis of the delocalization problem in the case of the
$1/R^{3}$ energy hopping amplitude \cite{b9}, the parameter $\chi _{\ast }$ is a 
decisive
parameter for the delocalization. As follows from Eqs. (\ref%
{eq:sh_dist_alt_res}) and (\ref{eq:l_dist_alt_res}), when $\chi _{\ast }\ll
1 $ the first resonance occurs at the distance $\ R_{a}\sim r_{\ast
}e^{1/\chi _{\ast }}\gg $ $r_{\ast }$. Thus, the first hop of the excitation
occurs to the distance $R_{a}$, taking some time $t_{1}$ determined by the
inverse hopping amplitude $t_{1}\sim R_{a}/U\propto e^{3/\chi _{\ast }}$.
Then, the time required for the second hop, when the next resonance appears,
is exponentially large compared with $t_{1}$\cite{b9}. Therefore, the
delocalization is exponentially slow, if any.

Consider the opposite case $\chi_{\ast }>1$. Then, the number of resonances
exceeds unity in each spherical layer $r_{1}<R<2r_{1}$ ($r_{1}>r_{\ast }$)
and the resonant sites form \textit{an infinite cluster} meaning an existence of the
delocalized state \cite{b7}. Thus, when a ratio of the external field
amplitude to the frequency is sufficiently large to provide the condition 
\begin{equation}
\chi _{\ast }>1, 
\label{eq:star1}
\end{equation}
the delocalization of excitations takes place.

The inverse time of a single hop between two nearest resonant neighbors in
the delocalization regime can be treated as a relaxation rate for the 
\textit{DE}
located at a certain site. The distance 
$R_{a}<r_{\ast}$ between these neighbors can be estimated from the relation 
$W(R_{a}) \approx 1$ (see Eq. (\ref{eq:sh_dist_alt_res})). This radius is determined as 
\begin{equation}
R_{a}\approx (Pa)^{-1/3}  \label{21a}
\end{equation}
and the typical dipole-dipole hopping amplitude corresponding to the distance 
$R_{a}$ is given by
Eq. (\ref{renampl1}) \vspace{-0.2cm} 
\begin{equation}
T\left( R_{a}\right) =\frac{U}{R_{a}^{3}}\sqrt{\omega /a}\approx \omega \chi
_{\ast }.  \label{rate}
\end{equation}
This expression gives an estimate of the inverse
lifetime  (the relaxation rate) for the \textit{DE} at an arbitrary site 
\vspace{-0.2cm}
\begin{equation}
\tau_{\ast }^{-1}\approx \omega \chi _{\ast }=\left( a\omega \right)
^{1/2}\chi  \label{yum}
\end{equation}

\section{Many body delocalization, dephasing and relaxation}

Here we will discuss the case of the zero field and a finite temperature 
$T>0$. 
The Anderson localization of excitations proved in Sect. 2 takes place 
only if every resonant pair \textit{RP} can be treated independently. This
assumption is valid provided that one can neglect the interaction between 
different excited  \textit{TLS}. We will see that this is not the case at any 
finite temperature $T>0$.  Below we will show
that at any finite temperature the long-range $1/R^{3}$ interaction of
excited  \textit{TLS} leads to the irreversible dynamics and 
relaxation. This relaxation is 
essentially 
of a many-body origin. In other
words, one should take into account simultaneous transitions in 
two or more \textit{RP} (see Fig. 3) and thus at least four  \textit{TLS} will 
participate in an
elementary process \cite{b7,iyp,b71,Burin 1989}.

Any \textit{RP} has four energy levels. Two of them
correspond to the flip-flop configuration mentioned above (see Fig. 2B). The two other 
states of a pair correspond to the configuration where both  \textit{TLS} are 
either
in their excited or their ground states. In fact, the typical energy of excited  
\textit{TLS} in 
a resonant pair is given by the thermal energy $T$, while their flip-flop 
transition 
amplitude Eq. (\ref{eq:transampl}) is generally much smaller due to the weakness 
of the interaction Eq. (\ref{eq:N0}). Therefore the 
flip-flop 
interaction $\Delta_{0p}$ can connect only pairs of  \textit{TLS},  
where one of them is in its excited state  
and the other one is in its ground state (Fig. 2). The other two states are 
separated from the flip-flop pair by the large energy gap of order of the 
temperature.

The states of the flip-flop configuration are separated by the energy
interval $\Delta_{p}=\mid E-E^{\prime}\mid$. In spite of the fact that 
$E,E^{\prime }\approx T$, one can construct \textit{RP} with 
$\Delta_{p},\Delta _{0p}\ll T$. Then, even if the interaction $V(R)$ between
these  \textit{TLS} is weak, the condition $\Delta_{p}\leq\Delta_{0p}$ can be 
valid.
Therefore the two levels in the resonant pair can be strongly coupled
(see Eq.(\ref{eq:reson})). In the remainder of this paper we will 
consider only flip-flop
configurations of \textit{RP}. We will treat \textit{RP} as a new
kind of the two-level system with the energy asymmetry $\Delta_{p}=\mid
E-E^{\prime}\mid$ and the tunneling amplitude $\Delta_{0p}(R)$ 
(\ref{eq:transampl}).

Thus below we investigate the relaxation of this novel \textit{RP} type of
the two-level system for which the distribution function of parameters 
$\Delta _{p}$ and $\Delta _{0p}$ is defined as \cite{b7,b6}
\begin{equation}
P^{(2)}(\Delta _{p},\Delta _{0p})=\langle \delta (\Delta _{p}-\mid
E-E^{\prime }\mid )\delta (\Delta _{0p}-\frac{U}{R^{3}}\frac{\Delta
_{0}\Delta _{0}^{\prime }}{EE^{\prime }})\rangle.  \label{k1}
\end{equation}
The brackets denote two averaging, namely, thermal averaging
and averaging over the distribution of parameters of original  \textit{TLS} 
(see Eq. (\ref{eq:AHWP})). In addition, the integration over the distance $R$ includes all possible pairs, so that the left-hand side of 
Eq.(\ref{k1}) can be rewritten as 
\begin{equation}
\int d^{3}R\int P(\Delta ,\Delta _{0})d\Delta d\Delta _{0}\int P(\Delta
^{\prime },\Delta _{0}^{\prime })d\Delta ^{\prime }d\Delta _{0}^{\prime
}n(E)\cdot
\end{equation}
\begin{equation}
(1-n(E^{\prime }))\delta (\Delta _{p}-\mid E-E^{\prime }\mid )\delta \left(
\Delta _{0p}-\frac{U}{R^{3}}\frac{\Delta _{0}\Delta _{0}^{\prime }}{%
EE^{\prime }}\right) ,  \label{k1'}
\end{equation}%
with $n(E)=[1+\exp (E/T)]^{-1}$ being the probability to find  \textit{TLS} in 
its
excited state. To evaluate Eq. (\ref{k1'}), we take
into account that the integral is determined
mainly by $E\sim E^{^{\prime }}\sim T.$ Therefore, due
to $\Delta _{p}\ll E\sim T$ one can omit $\Delta _{p}$ in the argument of
the first delta-function. Then one can easily evaluate the
pair distribution function Eq. (\ref{k1'}) within the logarithmic accuracy 
\cite{b7,b71} 
\begin{equation}
P^{(2)}(\Delta _{p},\Delta _{0p})\approx (PT)(PU)\frac{1}{\Delta _{0p}^{2}}.
\label{k2}
\end{equation}
Note that, the $\left( \Delta _{0p}\right) ^{-2}$ dependence
in the last expression is a consequence of the $R^{-3}$ interaction and
it is not related to the initial distribution Eq. (\ref{eq:AHWP}).
Although the distribution function Eq. (\ref{k2}) differs
from Eq. (\ref{eq:AHWP}), it also does not depend on an asymmetry 
parameter $\Delta_{p}$. The pair distribution function Eq. (\ref{k2}) has a stronger 
singularity at
small $\Delta_{0p}$ than the original TLS distribution function Eq.(\ref{eq:AHWP}) 
at small $\Delta_{0}$ . For this
reason, the concentration of low energy \textit{RP} excitations is
larger (and, correspondingly, the average distance between them is smaller)
than for initial  \textit{TLS}. The coupling constants $U(R)$ both for 
\textit{TLS} and
\textit{RP} are of the same order of magnitude, so that low energy \textit{RP}
interact stronger than \textit{TLS} (See Fig. 3). The reason for that 
is that having $N \gg 1$ \textit{TLS} we can make $N^{2} \gg N$ pairs of them.  

The ensemble of \textit{RP}, new kind of  \textit{TLS}, is described by the 
initial
Hamiltonian Eqs. (\ref{eq:spin}), (\ref{eq:int}). The only but a key
distinction from the initial model is that the distribution function (\ref
{k2}) should be used instead of Eq. (\ref{eq:AHWP}) at finite temperature $T>0$. 
Remember that resonant pairs are only those pairs for
which $\Delta_{p}\leq\Delta_{0p}$. Therefore each resonant pair has the single 
characteristic energy given by its flip-flop amplitude $\Delta_{0p}$. Since 
resonant pairs are mainly formed by  \textit{TLS} with  $\Delta \sim \Delta_{0} 
\sim T$, the charactreristic pair transition amplitude can be estimated as 
$U/R^{3}$ (\ref{eq:transampl}), where $R\sim (U/\Delta_{0p})^{1/3}$ is the size 
of the pair. 

Resonant pairs interact with each other. We will show following Refs. 
\cite{b7,b71} that this interaction inevitably leads to the energy 
delocalization induced by collective flip-flop transitions similar to that shown 
in Fig. 3. It is convenient to separate the whole sequence of resonant pairs 
into the infinite set of strips $k=1,2,3,..$. These strips are defined in the 
space of characteristic energies of resonant pair. Each strip $k$ is formed by 
resonant pairs having transition amplitudes within the range 
$(\Delta_{0p}(k)/2,\Delta_{0p}(k))$, with $\Delta_{0p}(k)=T/2^{k}$. One can 
estimate the concentration of \textit{RP} $N_{k}$ within the strip $k$ making 
use of the distribution function Eq. (\ref{k2})
\begin{equation}
N_{k}\approx (PT)(PU)=N_{\ast }.  \label{k4}
\end{equation}

\begin{figure}[tbp]
    \centering
    \includegraphics[width=5in,clip]{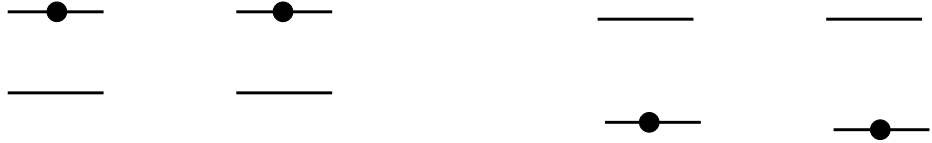}
    \caption{\label{Res2}
The interaction of resonant pairs leads to the many-body delocalization (A). The 
excitation is transferred due to the simultaneous flip-flop 
transitions of four  \textit{TLS} to their new states (B). The transitions involving two  \textit{TLS} 
belonging to different resonant pairs are not allowed because of the energy 
mismatch (C).}
\end{figure}

It is important that the \textit{RP} concentration in each strip  $k$ 
is completely independent of
the value $\Delta _{0p}(k)$. Separating strips corresponding to
all permissible values $\Delta _{0p}$, one can cover completely the whole
ensemble of resonant pairs. Within any strip the \textit{RP}
concentration is constant $N_{\ast }$ Eq. (\ref{k4}). Therefore the
average distance between \textit{RP} within any strip does not depend on
the kind of it. This distance is equal to $R_{\ast }\approx N_{\ast }^{-
\frac{1}{3}}$.
Since the interaction between \textit{RP} is of the same origin as
the interaction between \textit{TLS}, it is given by the same 
expression $U/R_{\ast}^{3}$ (see Eq.(\ref{eq:int})). Therefore, the typical energy of
the interaction between \textit{RP} within any strip is given by 
\begin{equation}
U_{\ast }\approx U/R_{\ast }^{3} \approx T\chi ^{2}.  \label{k5}
\end{equation}

Since \textit{RP} is a kind of  \textit{TLS}, one can introduce a concept of a
flip-flop configuration for two \textit{RP} shown in Fig. 3 and derive an
expression for
the transition amplitude $\Delta_{0}^{(4)}$ between the levels of this
configuration (which is similar to that of Eq.(\ref{eq:transampl})). Because 
interacting pairs are both resonant this transition amplitude is given by their 
characteristic interaction Eq. 
(\ref{k5}) (cf. Eq. (\ref{eq:transampl})). Consider the effect of this flip-flop 
interaction on \textit{RP} belonging to different strips. 

We should start with fastest, high energy resonant pairs of the strip 
$k=0$, 
while other pairs are slower and can be treated quasi-statically. It is clear 
that their characteristic interaction Eq. (\ref{k5}) is much 
smaller than their characteristic energy $T$ because of the weakness of the 
\textit{TLS} interaction $\chi \ll 1$ Eq. (\ref{eq:chi}). Thus,
they are isolated from each other. Their dynamics is faster than for other 
pairs having smaller energies, so they can be treated adiabatically with respect 
to other \textit{RP}. Then, the fast \textit{RP} possibly contribute to a small 
redefinition of parameters for slower pairs without any effect on their average
properties. Then to analyze the interaction effect for other pairs we should go 
all the way down in energy, increasing a strip index 
$k$. It is clear that the interaction becomes significant when the characteristic 
energy of pairs $\Delta_{0k}$ approaches the interaction energy Eq.(\ref{k5}). 
Remember that the interaction of \textit{RP} belonging to each strip is given by 
the universal value Eq. (\ref{k5}). The pairs with energy $\Delta_{0k}<U_{\ast}$ 
cannot be treated independently because their interaction is greater or equal to 
their characteristic energy. 

The strips with the energy smaller than $U_{\ast}$ satisfy the conditions of the 
Anderson delocalization because the characteristic flip-flop transition 
amplitude $U_{\ast}$ between \textit{RP} is greater or equal to their energy 
disordering, determined by their characteristic energy $\Delta_{0k}$. The 
elementary process of an energy hop between pairs is demonstrated in Fig. 3. It is 
important that in order to transfer the energy, the simultaneous 
transition of all four \textit{TLS} is 
required, while the transition of two \textit{TLS} belonging to different resonant pairs 
(Fig. 3C) is forbidden because of the energy mismatch of order of the thermal energy $T$.    
Thus \textit{RP} excitations in the cluster formed by the strip with the energy 
$U_{\ast}$ 
are delocalized. The relaxation rate of excitations within this strip
is given by the inverse characteristic interaction 
\begin{equation}
\tau_{\ast}^{-1}\sim U_{*} \approx T(PU)^{2}.
\label{k6}
\end{equation}

It turns out that the time $\tau_{*}$ defines the dephasing time within the 
ensemble of 
\textit{TLS}.
Consider the time-dependent fluctuation $\delta \Phi(t)$ of the phase at arbitrary probed  \textit{TLS}, induced by the energy-splitting 
fluctuation 
$\delta E(t)$. The proper dephasing time $\tau_{2}$ can be determined from
the relationship 
\begin{equation}
\delta \Phi (\tau_{2})=\tau _{2}\delta E(\tau _{2})\approx 1.  \label{k7}
\end{equation}
To show that the dephasing time is given by the time $\tau_{*}$, let us 
calculate the energy-splitting fluctuation $\delta E(t)$ 
within the time interval $t=\tau_{\ast }$. The overall density of  
\textit{RP} and, therefore, approximately the density of  \textit{TLS} undergoing 
the
transition during the time $t=\tau _{\ast }$ is of order of the \textit{RP} 
density $N_{\ast }$. Bearing in mind that every \textit{RP}
experiencing the transition at the distance $r$ from probed  \textit{TLS} 
contributes
a value $U(r)$ (see Eq. (\ref{eq:int})) into the energy-splitting fluctuation
for probed  \textit{TLS}, one can estimate 
\begin{equation}
\delta E(\tau _{\ast })=\sum\limits_{r}U(r)\approx N_{\ast }U\approx
U(R_{\ast }).
\label{eq:summ}
\end{equation}
This estimate was made taking into account that the sum in 
Eq. (\ref{eq:summ}) is defined by the largest term coming from 
the closest neighbour contribution.\cite{Halperin 1968} THis nearest neighbour 
is separated from \textit{TLS} under consideration by the distance 
$R_{\ast} = N_{\ast}^{-1/3}$.  

Then, taking into account Eq.~(\ref{k6}), one can find that $\tau _{\ast
}^{-1}\delta E(\tau _{\ast })\approx 1$. The comparison of this relation
with Eq. (\ref{k7}) allows us to link $\tau _{\ast }$ with $\tau _{2}$ and
therefore 
\begin{equation}
\tau _{2}^{-1}\approx U(R_{\ast })\approx T\chi ^{2}  \label{k8}
\end{equation}
This dephasing rate decreases linearly with the temperature. Compared with
the phonon-induced channel providing a $T^{2}$ -dependence (see Ref. 
\cite{Halperin 1968}), the dephasing governed by the mechanism concerned
predominates at sufficiently low temperatures.

Regardless of the relaxation mechanism, the spectral diffusion rate is
defined as \cite{Hunklinger,Halperin 1968} 
\begin{equation}
\left\langle \left\vert \dot{E}\right\vert \right\rangle \sim \tau
_{2}^{-2}.  \label{20a}
\end{equation}
On the other hand, in a system with the $1/R^{3}$ interaction this rate can be
expressed as the ratio of the spectral diffusion amplitude $\chi T$ and thermal
 \textit{TLS} relaxation time $\tau _{1}$ \cite{Hunklinger,Halperin 1968} 
\begin{equation}
\tau _{2}^{-2}=\chi T/\tau _{1}.  \label{halp}
\end{equation}
Eq. (\ref{halp}) is based on the time evolution of a TLS energy perturbed by 
transitions of neighboring \textit{TLS} in the system with $1/R^{3}$ 
interaction. If the characteristic relaxation time of  thermal TLS having the 
density $PT$  is given by $\tau_{1}$, then during the time $t<\tau_{1}$ the 
fraction of all TLS $n_{T}(t) \approx PTt/\tau_{1}$ undergoes transitions. The 
characteristic perturbation in each TLS energy caused by these 
transitions is defined as $Un_{T}(t) = UPTt/\tau_{1}$. The phase fluctuation can 
be estimated using Eq. (\ref{k7}) $\delta \Phi \sim UPT t^{2}/\tau_{1}$. 
Assuming that at $t=\tau_{2}$ the phase fluctuation becomes of order of unity 
(see Eq. (\ref{k7})) we derive Eq. (\ref{halp}). 
This derivation is correct, when TLS transitions have no correlations with each 
other. This is of course true for phonon stimulated relaxation, while in our 
case the spectral diffusion should be treated self-consistently. It was shown in 
the previous works \cite{b7,b71} that despite of certain memory effects in the 
spectral diffusion Eq. (\ref{halp}) remains approximately valid. 

Using (\ref{k8}) and (\ref{halp}) one can estimate the relaxation rate for
thermal  \textit{TLS} as \cite{b7} 
\begin{equation}
\tau _{1}^{-1}=\chi ^{3}T. 
\label{eq:linrel}
\end{equation}

\section{Relaxation at finite temperature and strong field}

Previously, we have shown that the $1/R^{3}$ 
interaction of excitations and the external alternating field can 
independently result in the 
delocalization and in the irreversible \textit{TLS} dynamics in the 
regime of arbitrary strong disordering. Under conditions of 
the real experiment both the \textit{TLS} interaction and the external 
alternating (measuring) field are significant. Below we consider the combined 
contribution of two those effects on the \textit{TLS} relaxation. 

Our analysis starts with the very low frequency of the external field, where the field 
can be treated as static. Then the frequency will substantially increase to 
its maximum value $\omega \sim T$. The case of larger frequences $\omega > T$ is 
beyond the scope of this paper because of almost unavoidable heating.  The 
results of our consideration are placed into the Table 1. 

\bigskip

Table 1. Frequency-dependent relaxation rate.

\begin{tabular}{|l|l|l|}
\hline
Frequency range & $\tau_{1}^{-1}$ & $\tau_{2}^{-1}$ \\ \hline\hline
$0<\omega<\left( T\chi^{2}\right) ^{2}/a$ & $T\chi^{3}$ & $T\chi^{2}$ \\ 
\hline
$\left( T\chi^{2}\right) ^{2}/a<\omega<a\chi^{2}$ & $\left( a\omega\right)
^{1/2}\chi$ & $\left( a\omega T^{2}\right) ^{1/4}\chi$ \\ \hline
$a\chi^{2}<\omega<T\chi$ & $a\chi^{2}$ & $\left( aT\right) ^{1/2}\chi^{3/2}$
\\ \hline
$T\chi<\omega<a$ & $\left( a/\omega\right) T\chi^{3}$ & $\left(
a/\omega\right) ^{1/2}T\chi^{2}$ \\ \hline
$a<\omega<T$ or $a<T\chi$ & $T\chi^{3}$ & $T\chi^{2}$ \\ \hline
\end{tabular}

\bigskip

Below we consider the case of a relatively large field amplitude $a > T\chi$ where 
the field effect is significant. One can show,   using similar methods as below, 
that for 
smaller field amplitudes $a < T\chi$ the field effect can be ignored and the 
relaxation is defined by the results of the previous section Eqs. (\ref{k8}), 
(\ref{eq:linrel}) for the relaxation and dephasing rates, 
respectively. It is interesting that the similar non-linearity criterion has 
been recently found in metallic glasses \cite{a>TPU}. 

Remember, that the parameter $T\chi= U(PT)$ determines the characteristic 
interaction of thermal \textit{TLS} ($\Delta_{0} \sim\Delta \sim T$). It 
represents the amplitude of the interaction induced spectral diffusion. The 
condition $a > T\chi$ means that the fluctuation of a TLS energy induced by the 
external field exceeds the interaction-induced fluctuation.   
We also assume that the field amplitude is less than the thermal energy to avoid 
heating. Thus in our consideration below the field amplitude $a$ satisfies the 
inequality 
\begin{equation}
T\chi < a < T. 
\label{eq:star}
\end{equation}

\subsection{Quasistatic field}

When the frequency of the external field $\omega $ is very small, the field is
almost static and we can ignore it. Then the relaxation and dephasing rates are 
defined by the zero field limit Eqs. (\ref{k8}), (\ref{eq:linrel}),  
respectively. In this regime the fluctuations of a
 \textit{TLS} energy are due to the self-consistent spectral diffusion alone. The 
spectral diffusion rate  is
given by Eq. (\ref{20a}). The rate of a  \textit{TLS}\ energy fluctuation 
induced by the external field is $a\omega$. The external field can be treated as 
static until the spectral diffusion
rate exceeds the fluctuation rate due to the field $\tau _{2}^{-2}>a\omega $%
, which takes place at $\omega <\left( T\chi ^{2}\right) ^{2}/a$ (see the first 
row in the Table 1). 

\subsection{Adiabatic relaxation induced by slow field}

At larger frequency  
\begin{equation}
\omega >(T\chi ^{2})^{2}/a
\end{equation}
the field cannot be ignored. It can stimulate transitions of TLS and 
irreversible relaxation  due
to energy level crossings of different  \textit{TLS}. At a large field amplitude 
Eq.(\ref{eq:star}) crossing between two  \textit{TLS} energy levels $E_{0} 
$ and $E_{j}$ ( $E_{0}-E_{j}-a\cos
\omega t=0$) \cite{amplitude} is possible provided that 
\begin{equation}
\left\vert E_{0}-E_{j}\right\vert <a.  \label{tlsres}
\end{equation}
The average distance between two  \textit{TLS} satisfying Eq. (\ref{tlsres}) is 
given
by $R_{a}\simeq \left( Pa\right) ^{-1/3}$ (see Eq. (\ref{21a})).

Consider a flip-flop pair of  \textit{TLS} with the close energies satisfying Eq. 
(\ref{tlsres}) and separated by the distance $R_{j}$. This pair undergoes level
crossing during the field oscillation period $\omega ^{-1}$. If the pair
transition amplitude is large $V^{2}=$ $\left( U/R_{j}^{3}\right) ^{2}>\dot{
E}\simeq a\omega ,$ the \textit{adiabatic relaxation} takes place.
Then according to the Landau - Zener theory the excitation will transfer within
a  \textit{TLS} pair with the almost \textit{unity probability} if the size 
$R_{j}$ 
of a pair is sufficiently small (see Eq. (\ref{cross_rad})) 
\begin{equation}
R_{j}<r_{\ast }\simeq \left( U/\sqrt{a\omega }\right) ^{1/3}.  \label{r*}
\end{equation}

The \textit{adiabatic} regime takes place when each thermal  \textit{TLS} has 
the
large number of adiabatic neighbors, satisfying Eqs. (\ref{tlsres}), (\ref{r*}). 
This requires $r_{\ast }>R_{a}$, which leads
to the frequency constraint from top (see Table 1, the second row) 
\begin{equation}
\omega <a\chi ^{2}  \label{case2}
\end{equation}%
For given thermal  \textit{TLS} the relaxation rate $\tau _{1}^{-1}$ can be 
estimated
as a frequency of adiabatic level crossings given by the product of the
field oscillation frequency $\omega $ and the number of adiabatic level 
crossing events during the single field oscillation. The latter number is given by the 
number of adiabatic neighbors $\chi_{\ast}$,
satisfying Eq. (\ref{tlsres}) and located within the sphere of the radius $%
r_{\ast }$ around given  \textit{TLS}, $\chi _{\ast }\simeq Par_{\ast 
}^{3}=\left(
a/\omega \right) ^{1/2}\chi \gg 1$ 
\begin{equation}
\tau _{1}^{-1}\simeq \chi _{\ast }\omega =\left( a\omega \right) ^{1/2}\chi
\label{t1-2001}
\end{equation}
It is interesting that the relaxation rate (\ref{t1-2001})
coincides with that of Eq. (\ref{yum}).

In the adiabatic regime we can estimate the
dephasing rate using Eqs. (\ref{halp}), (\ref{t1-2001}) 
\begin{equation}
\tau _{2}^{-1}=\left( T\chi /\tau _{1}\right) ^{1/2}=\left( a\omega
T^{2}\right) ^{1/4}\chi  \label{t2-2001}
\end{equation}
It is important that dephasing (\ref{t2-2001}) is so fast that the phase
coherency between the periodic events of level crossings occurring during
the period $\omega ^{-1}$ can be ignored. This is true in the whole
adiabatic frequency domain (the second row of the Table 1) because the
dephasing time is in fact less than the oscillation period (see Eqs. 
(\ref{eq:star}), (\ref{case2}), (\ref{t2-2001})) 
\begin{equation}
\omega \tau _{2}=(\omega /(a\chi ^{2}))^{3/4}(a/T)^{1/4}\chi ^{1/2}<1.
\label{low1}
\end{equation}

\subsection{Non-adiabatic regime at intermediate frequencies}

Consider the \textit{nonadiabatic regime}, which takes place at higher
frequencies 
\begin{equation}
\omega >a\chi ^{2}.  \label{20}
\end{equation}
This condition is opposite to Eq. (\ref{case2}). In this regime the vast
majority of thermal  \textit{TLS} undergo nonadiabatic level crossings (see Eq. 
(\ref
{nalc}) below).

In the adiabatic case we have dealt with the regime $\omega \tau _{2}<1$ (see
Eq. (\ref{low1})). In the nonadiabatic regime characterized by Eq.(\ref{20})
one should distinguish between two cases $\omega \tau _{2}<1$ and $\omega
\tau _{2}>1$ where the phase memory between two subsequent level crossings
either exists or does not exist, respectively.

Consider the first case $\omega \tau _{2}<1$ valid at the border of
adiabatic and non-adiabatic domains $\omega =a\chi ^{2}$ (see Table 1). Here
dephasing is fast and phase correlations between periodic level crossings
can be ignored. Most efficient flip-flop transitions occur between thermal
 \textit{TLS} separated by the average distance $R_{a}$ Eq. (\ref{21a}). In fact 
this distance is the characteristic separation of neighboring two-level systems 
with energy difference less than the field amplitude $a$, required for level 
crossing. On the other hand 
at larger distances a nonadiabatic transition probability decreases with the 
distance very fast as 
$V^{2}\sim R^{-6}$. Therefore the most efficient energy transfer should occur 
between nearest neighbors. The relevant transition amplitude is given by 
Eq. (\ref
{b32}) $V_{a}=U/R_{a}^{3}\sim a\chi $. In the large frequency case Eq. 
(\ref{20}) one has 
\begin{equation}
V_{a}^{2}<\dot{E}=a\omega  \label{nalc}
\end{equation}%
and transitions are \textit{nonadiabatic}. According to the Landau - Zener
theory, their probability per one energy level crossing, occurring a few times
for the period $\omega ^{-1}$, is $W_{a}\simeq V_{a}^{2}/\left( a\omega
\right) =\left( a/\omega \right) \chi ^{2}$. This probability 
defines the  \textit{TLS} relaxation rate as the inverse average time between 
two successful transitions
\begin{equation}
\tau _{1}^{-1}=W_{a}\omega \simeq a\chi ^{2}.  \label{17}
\end{equation}
Then, using Eq. (\ref{halp}), one finds 
\begin{equation}
\tau _{2}^{-1}=a^{1/2}T^{1/2}\chi ^{3/2}  \label{19}
\end{equation}%
The above derivation remains valid until breaking the condition Eq. (\ref
{low1}) at 
\begin{equation}
\omega >a^{1/2}T^{1/2}\chi ^{3/2}.  \label{45a}
\end{equation}

In the opposite case of $\omega \tau _{2}>1$, the field periodicity is
significant so that the formalism of dressed excitations (\textit{DE}) introduced in 
Sect.
3 (see also Ref. \cite{bkp2001}) becomes applicable.

The further analysis depends on the relationship between \textit{DE} energy 
splitting 
$\omega $ and the spectral diffusion amplitude $T\chi $ (see Eq. 
(\ref{eq:star})). 
We start with the case
of lower frequencies (see Table 1)
\begin{equation}
\omega \leq T\chi .  \label{3up}
\end{equation}%
where all pairs of  \textit{TLS} with an energy difference less than the field 
amplitude $a$ inevitably undergo \textit{DE} level crossing during the spectral diffusion
quasi-period $\tau _{1}$. These crossing levels are coupled by the
transition amplitude of dressed excitations Eq.(\ref{renampl1}).

In the regime of Eqs. (\ref{45a}), (\ref{3up})
the relaxation is induced by non-adiabatic level crossing caused by the
spectral diffusion. On the other hand, the spectral diffusion is caused by
the relaxation dynamics of  \textit{TLS}, so the process is self-consistent 
\cite%
{b7,iyp,b71,Burin 1989}.

Assume that there is an existing characteristic relaxation rate of thermal
 \textit{TLS} $\tau_{1}^{-1}$. Transitions of thermal  \textit{TLS} change other  
\textit{TLS} energies
giving rise to the spectral diffusion. The spectral diffusion leads to level
crossing of \textit{DE}, stimulating irreversible transitions with the output 
rate $
r_{out}$. The self-consistent relaxation mechanism requires the input rate $
\tau _{1}^{-1}$ to be equal to this output rate.

The rate of transitions induced by the spectral diffusion can be estimated
as the number of transitions induced by level crossings during the
quasi-period of the spectral diffusion $\tau _{1}$ multiplied by the
frequency of spectral diffusion cycles $\tau _{1}^{-1}$. Energy level crossing
for two \textit{DE} happens when (cf. Eq. (\ref{ander}), Fig. 2)
\begin{equation}
E_{1}-E_{2}=n\omega  \label{46a}
\end{equation}
(where $n$ is an integer number). The total number of such crossings due to the
spectral diffusion induced energy fluctuation $T\chi $ for the time $\tau
_{1}$ is given by 
\begin{equation}
N_{\tau _{1}}\simeq (T\chi /\omega ).  \label{6.1}
\end{equation}
Multiple crossings for the same  \textit{TLS} pair give a logarithmic correction 
to
Eq. (\ref{6.1}) and can be neglected in the qualitative scaling approximation \cite{b7,b71,Burin 1989}.
This is due to the specific behavior of the spectral diffusion induced by the
$1/R^{3}$ interaction.  For this spectral 
diffusion the characteristic energy fluctuation is directly proportional to the 
time \cite{Hunklinger,Halperin 1968}. Therefore this is the anomalous diffusion 
process compared to the 
normal diffusive $t^{1/2}$ behavior of the displacement. For this super-
diffusion case the probability of energy to return back to its initial value is 
not so large as in the case of the normal diffusion.  

We consider nonadiabatic flip-flop transitions between only neighboring thermal
\textit{DE} separated by the distance $R_{a}$ (see Eq. (\ref{21a})), because the 
nonadiabatic
transition probability to larger distances drops with the distance very fast ($R^{-6}$). The probability of the nonadiabatic transition during one level
crossing for such a pair of thermal \textit{DE} induced by the spectral 
diffusion
with the rate $<\mid \dot{\varepsilon} \mid >\approx \tau _{2}^{-2}$ can be
estimated using the Landau-Zener theory in the nonadiabatic limit. The
transition probability can be found using the transition amplitude
Eq. (\ref{renampl1}) for $R=R_{a}$ 
\begin{equation}
W_{\ast }=\left( V_{a}\tau _{2}\right) ^{2}<1  \label{nonadia}
\end{equation}
The relaxation rate is expressed as 
\begin{equation}
\tau _{1}^{-1}=r_{out}=N_{\tau _{1}}W_{\ast }/\tau _{1}\simeq a\chi ^{2}.
\label{eq:tr_rate}
\end{equation}
One can show that if we take the input relaxation rate slower than 
Eq. (\ref{eq:tr_rate}), the output rate will be faster than the input one, and, if we
take the input rate larger than Eq. (\ref{eq:tr_rate}) the output rate will be smaller than the input rate. Therefore the solution (\ref{eq:tr_rate}) is stable and the only
possible. Thus, the \textit{nonadiabatic single particle} relaxation
described by Eqs. (\ref{17}), (\ref{19}) takes place within the whole
frequency domain $a\chi^{2}<\omega <T\chi $ (Table 1, the third row).

\subsection{High frequencies}

When the external field frequency $\omega $ exceeds the scale $T\chi$, the
spectral diffusion does not necessarily lead to \textit{DE} energy level crossing 
for closely located pairs, considered
in Sect. 5.3. The single-particle delocalization in this case
does not occur similarly to the case studied in Sect. 2 because the field frequency is 
large compared to 
that needed for the delocalization (see Table 1, line 2).  On the other hand, 
there exists a finite concentration of excited \textit{DE}
(excited  \textit{TLS} dressed by some certain number of the external field quanta, 
see Fig. 2) 
and many-body relaxation should be similar to that described in Sect. 4.

Following the approach of Sect. 4  one can introduce the
concept of the resonant pair of \textit{DE}, instead of \textit{RP} (resonant 
pairs of  \textit{TLS} ).
A resonant pair of \textit{DE} (\textit{RPDE}) is defined as a pair of 
\textit{DE} separated by the
distance $R_{12}$ with sublevel energies $E_{1}$ and $E_{2}$, which 
obeys the resonant condition for some integer $n<a/\omega$ (see Fig. 2)
\begin{equation}
\Delta_{pn}=\left\vert E_{1}-E_{2}+n\omega \right\vert <T\left( R\right).
\label{rpderc}
\end{equation}
Here the transition amplitude $T\left( R\right) \approx \frac{U}{R^{3}}\sqrt{
\frac{\omega }{a}}$ is taken from Eq. (\ref{renampl1}). Like a resonant
pair, one can treat \textit{RPDE} as a new kind of two-level system. The 
parameter $\Delta_{pn}$ is the asymmetry energy for \textit{RPDE}. Then,
strictly following the derivation between Eq. (\ref{k1}) and Eq. (\ref{k4})
one can estimate the concentration of \textit{RPDE} for any given value of
transition amplitude $T\left( R\right)$. This concentration is given by the
expression 
\begin{equation}
N^{\ast }\approx T\chi ^{2}\sqrt{\frac{a}{\omega }}  \label{rpdeconc}
\end{equation}
The concentration $N^{\ast }$ is independent of the given parameter 
$T(R)$
just like the concentration of \textit{RP} $N_{\ast }$ defined by Eq. (\ref{k4}) 
does not
depend on the parameter $\Delta _{0p}$. The appearance of the factor $\sqrt{
\frac{a}{\omega }}$ in Eq. (\ref{rpdeconc}) in comparison with Eq. (\ref{k4})
can be understood as follows. The concentration $N^{\ast }$ compared with $
N_{\ast }$ acquires the factor $n_{\ast }=a/\omega $ since resonance
condition (\ref{rpderc}) should be valid for at least one integer $\left\vert
n\right\vert <n_{\ast }$ so that the number of possible distinguishable resonances 
for the single pair multiplies the resonance probability by the factor 
$n_{*}$. On the other hand, the resonance probability  decreases by the factor 
$1/\sqrt{n_{\ast }}$ proportionally to the reduction in the transition
amplitude Eq. (\ref{renampl1}). The overall effect is just the increase in the 
density of
resonant pairs by the factor $\sqrt{a/\omega}$ (cf. 
Eq.(\ref{eq:l_dist_alt_res})). 
Accordingly, the
interaction between these resonant \textit{DE} pairs enhances by the same 
factor and
the new dephasing rate is given by this universal interaction
\begin{equation}
1/\tau _{2}=T\chi ^{2}(a/\omega )^{1/2}.  \label{eq:dephasing4}
\end{equation}%
Making use of Eq.(\ref{halp}) one can estimate the relaxation rate as 
\begin{equation}
\tau _{1}^{-1}=\left( T\chi \tau _{2}^{2}\right) ^{-1}=(a/\omega )T\chi ^{3}.
\label{T1RPD}
\end{equation}

When the frequency exceeds the amplitude $a < \omega<T$, excitation dressing
becomes negligibly small and we return to the linear regime (the last row of
the Table 1). The regime when the frequency exceeds the temperature is realized
in several echo measurements \cite{b11-01}. This regime can lead to heating and requires
a special study.

\section{Discussion}

We have described the self-relaxation rate of  \textit{TLS} at different 
temperatures,
external field amplitudes, and frequencies. We found that the relaxation
rate is either temperature independent (see the Table 1), or decreases with
the temperature as $T$. The phonon stimulated relaxation $\tau
_{1}^{-1}\propto T^{3}$ (see Ref. \cite{Hunklinger}) is much slower in the low
temperature limit and the self-relaxation described above should dominate
when $T\rightarrow 0$.

The suggested theory predicts the parametric dependences of relaxation rates in 
various regimes. In  contrast to the theory for the phonon-stimulated 
relaxation (see the review of Hunklinger and Raychaudchary \cite{Hunklinger} and 
references therein) we are not able to determine the numerical coefficients for 
each rate process shown in the Table 1. This missed factor can 
be of order of unity, but it can also be much greater or much smaller than unity, 
like $0.01$ or $10^{2}$. The example of the latter situation is given by the 
tunneling rate involving two-phonon processes, examined by Kagan and Maksimov 
\cite{classics1} for the quantum diffusion problem, where they found the large 
numerical prefactor of $1000$ due to the large factorial factor involved. 

We do also expect large numerical prefactors for our expressions. There are 
several reasons to have them. First, each factor $\chi$ possibly involves the 
spherical integration factor $4\pi$ or 
$4\pi/3$ in addition to $PU$ factor. In this paper we have included the factor 
$4\pi$ into the definition of our factor $\chi$ Eq. (\ref{eq:chi}) contrary to 
the previous work. This is done because this factor naturally appears in 
the localization criterion (\ref{eq:N0}) and correspondingly it can enter the rate 
expressions of the table 1.  Thus the parameter of interaction weakness $PU \sim 
10^{-4} - 10^{-3}$ can be increased  by the order of magnitude. Accordingly, the 
``linear" expression for the relaxation rate Eq. (\ref{eq:linrel}) agrees better with  the results of the systematic internal friction measurements performed by 
Classen, Burkert, Enss, and Hunklinger \cite{b11-2}. We believe that the factor 
$\chi=4\pi PU$ reflects the absolute values of relaxation rates better than the 
smaller factor $\chi=PU$ itself. The accurate calculations of the numerical factors 
are beyond the scope of our qualitative study.  The alternative explanation 
for a quantitative disagreement 
\cite{b11-2} can be based on the strong non-linearity. For instance, the extra-
factor $a/\omega$  in the relaxation rate (see Eq. (\ref{T1RPD})) can account 
for the difference of theory and experiment as well. The comparison of our 
predictions with experiment can still be performed using the experimental 
data for different glasses, with different values of the parameter $\chi$.  
Note that the preliminary numerical analysis of the Floquet state 
delocalization, described in Sect. 3, in the equivalent one-dimensional model 
supports our expectations of the large numerical prefactor, exceeding unity by 
at least one order of magnitude. 

Another possible problem of the direct application of our theory to the 
experiment is that the interaction-induced relaxation leads to the equilibration 
within the TLS subsystem rather than for the whole system. This is similar to spin-
spin and spin-lattice relaxations in the NMR problem. Then in order to to describe the heat 
balance between TLS and phonons, one should introduce the separate TLS and phonon 
temperatures and perform the thermal balance analysis for the whole system. This 
study is beyond the scope of this paper.  
 
The crossover temperature between two regimes depends on the external field
parameters $a$ and $\omega$. It usually reduces with decreasing the field
amplitude and can also decrease with decreasing the field frequency (see the
Table 1). This knowledge helps us to understand the absence of the interaction-stimulated relaxation reported by Pohl and coworkers in Ref. \cite{Pohl}. In
this work, the strain field amplitude was made extremely small $\varepsilon
\leq10^{-8}$, while the frequency $\omega\sim0.5MHz$ is higher than in other
group studies \cite{b11,b11-1,b11-2} and belongs to the range described by
the fourth column of the table 1, when the relaxation rate decreases with
increasing $\omega$.

Our results agree with the recent low-temperature measurements of the
dielectric constant by Ladieu and coworkers \cite{m6,ladieu1}, which can be
interpreted assuming that the  \textit{TLS} relaxation rate becomes temperature
independent. This shows the dependence on the external field amplitude $%
\sqrt{a}$ (the second row of the Table 1). Another very interesting discovery
of Ladieu and coworkers includes the sensitivity of  the \textit{TLS} relaxation
rate to the sample thickness. The suggested theory is essentially three
dimensional. If the thickness of the sample will be less than the typical
distance between resonant pairs (around few tens of nanometers) then the
irreversible relaxation will remarkably slow down. This agrees with the
observations of Ref. \cite{m6,ladieu1}. More accurate theoretical analysis of
the data is necessary for the careful interpretation.  

Our predictions can be directly verified using the non-equilibrium measurement
technique developed by Osheroff and coworkers \cite{b11-1,osher2003-1},
which is based on the analysis of the system response to the large sweep of the
external electric field $E_{DC}$ taking different times $\tau_{s}$. Our
results for the field-stimulated relaxation are valid in this case if one
takes $a\sim E_{DC}\mu_{TLS}$ (where $\mu_{TLS}\sim 1D$ is the typical 
\textit{TLS} dipole moment) and $\omega\simeq1/\tau_{s}$. Since the maximum
possible amplitude $a$ is very high $a \geq 0.1 K$, while the minimum
``frequency" is as low as $\omega\sim1s^{-1}$, all the regimes described in the
Table 1 can be attained and analyzed.

\section{Acknowledgment}

The work of A. L. Burin is supported by TAMS GL fund no. 211043 through the
Tulane University. The work of I. Ya. Polishchuk is supported by Grant RFBR
2004. We are grateful to organizers and participants of the International
Workshop \textquotedblleft Collective phenomena in glasses at low
temperatures" (Dresden, 2003) for many useful discussions and suggestions. We
also wish to thank  Christian Enss, Peter Fulde, Yuri Kagan, Leonid Maksimov, 
Douglas Osheroff, Niina Ronkainen-Matsumo and Boris Shklovskii for valuable 
discussions, suggestions and ideas.

Looking at the Reference list, the reader can easily ascertain that it
is hard  to overestimate the contribution of Professor Siegfried Hunklinger
to the investigation of glasses at low temperature. The authors have many ideas, 
suggestions and advises  from the discussions with him. We are
glad to have an opportunity to dedicate this paper to his 65th birthday. It
is especially pleasing for us that this work develops the theory of the 
\textit{TLS}
interaction discovered in Professor Siegfried Hunklinger's  early researches \cite{hun1}.

\section{Appendix}

To solve Eq. (\ref{efse2}), let us introduce the partial amplitudes of the
Floquet states $c_{in}$ so that
\begin{equation}
b_{i}=\exp \left[ -i\varepsilon t\right] \sum_{n}c_{in}\exp \left( -in\omega
t\right)
\end{equation}
Substituting this expression into Eq. (\ref{efse2}) and taking its Fourier
transform, we arrive at the equation 
\begin{eqnarray}
\varepsilon c_{im} =\left( E_{i}-m\omega \right) c_{im}-\frac{a_{i}}{2}
\left( c_{i\,m+1}+c_{i\,m-1}\right)  \label{transform} 
\nonumber\\
 +\sum_{j}U_{ij}c_{jm}. 
\end{eqnarray}
Then multiply the last equation by the Bessel function $J_{n-m}\left( \frac{
a_{i}}{\omega }\right) $ and perform the summation over $m.$ The result can
be presented in the form 
\begin{equation}
\begin{array}{l}
\varepsilon \sum_{m}c_{im}J_{n-m}\left( \frac{a_{i}}{\omega }\right) =\left(
E_{i}-n\omega \right) \sum_{m}c_{im}J_{n-m}\left( \frac{a_{i}}{\omega }%
\right) + \\ 
+\sum_{jm}U_{ij}c_{jm}J_{n-m}\left( \frac{a_{i}}{\omega }\right)
+\sum_{m}c_{im}\left( \left( n-m\right) \omega \cdot \right. \\ 
\left. \cdot J_{n-m}\left( \frac{a_{i}}{\omega }\right) -\frac{a_{i}}{2}%
\left( J_{\left( n-m\right) +1}\left( \frac{a_{i}}{\omega }\right)
+J_{\left( n-m\right) +1}\left( \frac{a_{i}}{\omega }\right) \right) \right)%
\end{array}
\label{bes1}
\end{equation}
Due to the recurrsion properties of the Bessel functions, each term in the last
sum of the above equation vanishes. Finally, using the Graph
summation formulae 
\begin{equation}
J_{n-m}\left( \frac{a_{i}}{\omega}\right) =\sum_{p}J_{p-n}\left( \frac{a_{j}%
}{\omega}-\frac{a_{i}}{\omega}\right) J_{p-m}\left( \frac{a_{j}}{\omega}%
\right) .
\end{equation}
in the sum $\sum_{jm}$ in Eq. (\ref{bes1}) and introducing the notation 
\begin{equation}
d_{in}=\sum_{m}c_{im}J_{n-m}\left( \frac{a_{i}}{\omega}\right)
\end{equation}
we obtain 
\begin{eqnarray}
\varepsilon d_{in}=\left( e_{i}-n\omega\right) d_{in}+\sum_{jp}\tilde{U}%
_{in;jp}d_{jp},  
\nonumber\\
\tilde{U}_{in;jp}=U_{ij}J_{p-n}\left( \frac{a_{j}-a_{i}}{
\omega}\right)  \label{mae}
\end{eqnarray}


\end{document}